\documentclass[12pt]{article}

\newtheorem{Theorem}{Theorem}[section]
\newtheorem{Definition}[Theorem]{Definition}
\newtheorem{Proposition}[Theorem]{Proposition}
\newtheorem{Lemma}[Theorem]{Lemma}

\makeatletter
\usepackage{latexsym}
\usepackage[T1]{fontenc}
\usepackage{amsmath}
\usepackage{amssymb}

\newtheorem{Corollary}[Theorem]{Corollary}

\def \AO {{\cal A}({\cal O})}
\def \AO' {{\cal A}({\cal O}')}

\def \] {\supseteq}

\def \Pf {{\bf Proof.\,\,}}

\def \be {\begin{equation}}
\def \ee {\end{equation}}

\def \ume {{\scriptstyle{\frac{1}{2}}}}
\def \ra {\rightarrow}

\def \eqq {\equiv}

\def \a {{\alpha}}

\def \g {{\gamma}}
\def \d {{\delta}}
\def \eps {{\varepsilon}}

\def \l {{\lambda}}

\def \A {{\cal A}}
\def \B {{\cal B}}
\def \C {{\cal C}}

\def \G {{\cal G}}
\def \H {\mbox{${\cal H}$}}

\def \L {{\cal L}}

\def \O {{\cal O}}

\def \S {{\cal S}}

\def \U {{\cal U}}

\def \W {{\cal W}}
\def \Z {{\cal Z}}

\def \id {{\bf 1 }}

\def \Rbf {{\bf R}}

\def \AO {{\cal A}({\cal O})}
\def \AO' {{\cal A}({\cal O}')}

\def \] {\supseteq}

\def \Pf {{\bf Proof.\,\,}}

\newfam\Ssfam
\font\eleSs=cmss10 at12pt \font\sevenSs= cmss10 at 8pt \font\sixSs= cmss10 at 6pt

\textfont\Ssfam=\eleSs\scriptfont\Ssfam=\sevenSs%
\scriptscriptfont\Ssfam=\sixSs \def\Ss{\fam\Ssfam\eleSs}

\def\doppio#1{{\rm I}\kern-.1667em{\rm #1}}

\def\Q{\text{Q}\kern-.52em
    \text{\vrule height1.5ex width.5pt depth0pt}\kern.45em}

\def\Z{{\mathchoice {\hbox{$\Ss\textstyle Z\kern-0.4em Z$}}
{\hbox{$\Ss\textstyle Z\kern-0.4em Z$}} {\hbox{$\Ss\scriptstyle Z\kern-0.25em
Z$}} {\hbox{$\Ss\scriptscriptstyle Z\kern-0.2em Z$}}}}

\def\C{{\mathchoice{\hbox{$\rm\textstyle\text{\kern.35em\vrule
   height1.5ex width.5pt depth0pt\kern-.35em C}$}}
{\hbox{$\rm\textstyle\text{\kern.35em\vrule
   height1.5ex width.5pt depth0pt\kern-.35em C}$}}
{\hbox{$\rm\scriptstyle\text{\kern.35em\vrule
   height1.5ex width.3pt depth0pt\kern-.35em C}$}}
{\hbox{$\rm\scriptscriptstyle\text{\kern.35em\vrule
   height1.5ex width.2pt depth0pt\kern-.35em C}$}}}}

\def \be{\begin{equation} \displaystyle}
\def \ee{\end{equation}}

\def \A*{\mbox{$A^{*} $}}
\def \B*{\mbox{$B^{*} $}}
\def \C*{\mbox{$C^{*} $}}

\def \id{\mbox{${\bf 1}\,$}}

\def \bea{\begin{eqnarray}}
\def \eea{\end{eqnarray}}

\def \Pf{{\em Proof.\,\,}}

\def \a{\alpha}

\def \g{\gamma}
\def \l{\lambda}
\def \d{\delta}

\@addtoreset{equation}{section}

\def \be {\begin{equation} \displaystyle}

\def \ee {\end{equation}}

\def \ra {\rightarrow}
\def\AO {\mbox{${\cal A}({\cal O})$}}
\def\AO'{\mbox{${\cal A}({\cal O}')$}}
\def\O {\mbox{${\cal O}$}}

\def\A{\mbox{${\cal A}$}}

\def \ra{\rightarrow}

\def \eps {{\varepsilon}}

\def \O {{\cal O}}

\def \A {{\cal A}}
\def \AO {\A(\O)}
\def \AOl'{\A(\O_{loc}')}

\def \B {{\cal B}}

\def \H {{\cal H}}

\def \S {{\cal S}}

\def \gm {{g_m}}

\def \g {{\gamma}}

\def \M {{\cal M}}

\begin{document}
\begin{titlepage}
\title{Quantum Mechanics on Manifolds and Topological Effects \thanks{A
preliminary version of this work was presented as a joint work by the first
author at the Workshop "Local Quantum Theory", Vienna Sept. 1997, and made
available to the participants as  hand written notes }}

\sloppy

\author{G. Morchio \\Dipartimento di Fisica, Universit\`a di Pisa,
\\and  INFN, Sezione di Pisa, Pisa, Italy \and F. Strocchi \\
Scuola Normale Superiore, Pisa, Italy \\ and INFN, Sezione di Pisa}

\fussy

\date{}

\maketitle

\begin{abstract}

A unique classification of the topological effects associated to quantum
mechanics on manifolds is obtained on the basis of the invariance under
diffeomorphisms  and the realization of the Lie-Rinehart relations between the
generators of the diffeomorphism group and the algebra of $C^\infty$ functions
on the manifold. This leads to a unique (\lq\lq Lie-Rinehart\rq\rq )
$C^*$-algebra as
observable algebra; its regular representations are shown to be locally
Schroedinger and in one to one correspondence with the unitary representations
of the fundamental group of the manifold. Therefore, in the absence of
spin degrees of freedom and external fields, $ \pi_1(\M) $
appears as the only source of topological effects.

\end{abstract}

\vspace{15mm} \noindent
Math. Sub. Class.: 81Q70, 81R15, 81R10

\vspace{3mm}  \noindent

\end{titlepage}

\newpage
\section{Introduction}
The standard formulation of Quantum mechanics (QM) is based on canonical
quantization and its foundational problems have been clarified in terms of the
identification of observable algebras and the classification of the
corresponding states. A full control of their structure has been obtained by
the identification of the observable algebra with the (unique $C^*$-)algebra
generated by the exponentials of the canonical Heisenberg variables $q_i, p_i$
and by the  uniqueness of its Hilbert space (regular) representation. The
extension of such a strategy to a formulation of QM on manifolds still presents
substantial open problems. The basic issue is the identification of the
observable algebra playing the role of the Weyl algebra and in fact different
choices have led to different mathematical structures and different physical
results.

\def \cifm {C^\infty(\M)}
\def \difm  { \mbox{Diff}(\M)}
\def \gm {\G(\M)}
\def \gtm {\tilde{\G}(\M)}
\def \cmo {{C_0^\infty(\M)}}
\def \cif {{C^\infty}}
\def \co   {{C^\infty_0(\O)}}

The first important result in this direction is due to Segal ~\cite{Se}, who
emphasized the role of the group Diff$(\M)$ of diffeomorphisms of the manifold
$\M$ and its action on the algebra $C^\infty(\M)$ of $C^\infty$ functions on
$\M$. His strategy can be read as the identification of the observable
$C^*$-algebra as  the crossed product $\cifm \times \difm$ and his results
essentially amount to a classification of its (regular)  unitary
representations, which satisfy the crucial additional assumption that $\cifm$
is represented by a maximal abelian subalgebra. Actually, under this assumption
one has a unique (Schroedinger) representation, apart from possible phase
factors arising from the one dimensional cohomology of the manifold, leading to
a very restricted class of topological effects.

The problem of QM on manifolds was rediscussed by Landsman~\cite{La}, with the
aim of a systematic analysis of topological effects. Following
Mackey~\cite{Ma}, Landsman considered quantum mechanical systems whose
configuration space $Q$ can be represented as a homogeneous space $G/H$, with
$G$ a locally compact group, $H$ a subgroup of $G$ and identified  the
observable $C^*$-algebra with the crossed product $G/H \times G$. In
particular, if $Q$ is a manifold $\M$, $G$ can be a Lie subgroup of $\difm$;
then $H$ is the stability group of one (arbitrary) point of $\M$ and the
generators of $G$ play the role of momenta.

\def \cst   {{C^*}}
For a given $\M$, the so obtained QM crucially depends on $G$; in fact,
different choices of $G$ lead to different $C^*$-algebras of observables and
also to different topological effects. Furthermore, contrary to Segal's
approach, one does not have invariance under diffeomorphisms (since $\difm$ is
not locally compact and does not have invariant proper subgroups~\cite{Mi}).

Segal's strategy of classifying representations  of the entire $\difm$ has been
reproposed by Doebner et al.~\cite{Do}, who generalized Segal's analysis by
relaxing the maximality of $\cifm$ and by allowing deviations from the Lie
algebra relations of $\difm$, corresponding to the introduction of a (gauge)
connection. On the other side, they restricted  their analysis to the
representations of the vector fields which generate $\difm$ in an $L^2$ space
of sections of a vector bundle over the manifold, with finite dimensional
fibers, with suitable differentiability properties. A classification of such
representations in terms of topological effects is obtained under additional
simplifying assumptions, with results which only partially compare with those
by Landsman.

\def \pium {{\pi_1(\M)}}

The occurrence of quantum topological effects has been investigated also within
approaches to QM in terms of classical trajectories (path integral and Bohmian
mechanics)~\cite{Du}, naturally leading to a classification in terms of the
fundamental group $\pi_1(\M)$ of the manifold.

The restrictions  and choices underlying the above analyses, while technically
effective in view of the resulting classifications, leave open the question of
the validity and derivation of the so obtained effects on the ground of general
basic principles. The aim of this note is to provide a formulation of QM on
manifolds exclusively based on the identification of a unique observable
algebra from fundamental (physical) principles (excluding for the moment spin
degrees of freedom and  external fields):

\noindent 1) ({\bf Localization and Lie relations}.) The observable algebra
should be generated  by localized variables playing the role of positions and
momenta. This leads to choose  as position observables  $\cmo$, the $\cif$
functions on $\M$ of compact support, and, as  "momentum" observables,
variables $T_v$ indexed by a  Lie algebra of vector fields $v$ of compact
support, reproducing the linear and  Lie algebra relations between the vector
fields $v$ and their action as derivations on $\cmo$.

\def \lm {{\cal L}(\M)}
\noindent
2) ({\bf Diffeomorphism invariance}.) The identification of the
observable algebra should be independent of any choice of coordinates or of
additional geometrical constraints, i.e. it should be diffeomorphism invariant.
This leads to take the Lie algebra $\lm$ of {\em all} the $\cif$ vector fields
of compact support; this is actually the only possibility for compact $\M$,
since there is no diffeomorphism invariant subalgebra of $\lm$.

It is important that, without additional qualifications, different vectors
fields are treated as independent and therefore, even at the local level, one
has no relation between the number of independent momenta and the dimension of
the manifold, as one would expect on physical grounds. The point is that the
Lie relations between $\difm$ and $\cifm$ are a too general mathematical
structure and their interpretation is far from unique; in particular the same
Lie relations appear in the description of all $N$ particle systems on the same
manifold $\M$, as the non-relativistic local current algebra relations
~\cite{Goldin}. Moreover, a general classification of the representations of
$\cifm \times \difm$ is a difficult open mathematical problem. In fact, the
additional {\em ad hoc} requirements introduced in the literature (maximality
of $\cifm$ by Segal, fiber bundle restrictions by Doebner et al. and
restrictions to subgroups of $\difm$ by Landsman) have the purpose and the
effect of simplifying   the mathematical problem through an elimination of
unwanted degrees of freedom; however, as discussed above, the so derived
quantum mechanical effects substantially depend on such optional choices. A way
out of such unwanted degrees of freedom, compatibly with the above requirements
1,2, is obtained by taking into account the dependence relations between vector
fields through multiplication by $\cmo$, i.e. by realizing that the Lie algebra
$\lm$ of vector fields is a (diffeomorphism invariant) module over $\cmo$.
Actually, it is enough to realize such dependence relations at the strictly
local level, i.e. for vector fields and $\cif$ functions with supports in
regions $\O$ diffeomorphic to open spheres. This argument leads to

\def \tv {T_v}
\def \lo {\L(\O)}
\def \ulv {U(\l v)}
\def \rv {R_v}
\def \rlv {R_{\l v}}
\def \rav {{R_{\a v}}}
\def \am {\A(\M)}

\noindent
3) ({\bf Elimination of redundant degrees of freedom}.) The momenta
$T_v$, indexed by vector fields $v$, satisfy the following algebraic relations
\be{T_{f v} = \ume ( f \tv + \tv f),\,\, \,\,\,\forall f \in \co, \,\,\,\forall
v \in \L(\O).}\ee
Mathematically, the algebraic operations in the right hand
side of eq.\,(1.10 is assumed to reproduce the intrinsic Lie-Rinehart product
$\co \times \lo \ra \lo$ which makes $\lo$ a Lie-Rinehart (LR) algebra over
$\co$ ~\cite{LR}.

In order to define a $C^*$-algebra of observables on the basis of 1-3 we take
as generators the algebra $\cmo$, the one parameter groups $U(\l v)$, and the
resolvents $R_v$, $ v \in \lo$, (of the corresponding generators $\tv$), in
terms of which condition (1.1) can be imposed. In this way we shall obtain a
unique $\cst$-algebra $\A(\M)$ (Sect.\,2).

The module structure of $\lm$ on $\M$, together with its vector bundle
structure on $\M$, gives rise to a Lie algebroid and the relevance of this
geometric structure for the problem of quantization of Poisson manifolds has
been discussed in the literature (see ~\cite{Hue}, ~\cite{Lan}). However, the
vector bundle structure cannot be shared by the quantum observable algebra and
the standard crossed product $C^*$ structure associated to  the corresponding
Lie groupoid is not enough, since it does not include and does not imply the
product relations (1.1) for the generators of the Lie groupoid. On the other
hand, as discussed above, the LR relations (1.1) and their local structure are
essential ingredients for the very identification of the observable
$C^*$-algebra and for a unique classification of the topological effects
associated to QM on manifolds. With respect to the crossed product
$C^*$-algebra, $\A(\M)$ is therefore a better candidate for a $C^*$-algebraic
non-commutative version of the cotangent bundle on $\M$ with its symplectic
structure.

The main result of this note is the classification of all the representations
of $\am$, in which $\ulv$ are strongly continuous in $\l$ and the generators
satisfy the Lie algebra relations on a dense invariant domain ({\em regular
representations}). All such representations will be shown to be locally, i.e.
for regions $\O$ as above, unitarily  equivalent, apart from multiplicities,
to the Schroedinger representation
in $ \H \eqq L^2(\M, d \mu)$, in which $\cmo$ act
as multiplication operators and $\forall \psi \in \H$, $g \in \difm$
\be{U(g)\psi(x) =
\psi(g^{-1} x) J(g, x), \,\,\,\,J(g,x) \eqq [d \mu(g^{-1} x)/
d \mu(x)]^{1/2}  }\ee with $ d \mu$ absolutely continuous with
respect  to the Lebesque measure (Sect.\,3).

Globally, as a consequence of the elimination at the local level of the
redundant degrees of freedom (Lie-Rinehart  $C^*$-algebra), the regular
representations are in one to one correspondence with the unitary irreducible
representations of $\pium$, the first homotopy group of $\M$, which thus
appears as the only source of topological effects, in the absence of
additional locally observable degrees of freedom and external fields
modifying the Lie product of $\lm$.
In particular, for simply connected manifolds one has a uniqueness
theorem as the Von Neumann theorem for the Weyl algebra (Sect.\,4).

All the regular representations of $\am$ can be realized as Schroedinger
representations on  functions on  the universal covering space of $\M$,
yielding unitary representations of $\pium$. It is worthwhile to remark that
the role of the universal covering space of $\M$ here emerges from first
principles, rather than from the somewhat arbitrary classification of the
\lq\lq  classical trajectories\rq\rq\ in the functional integral formulation.
The intrinsic {\em a priori} topological structure is in fact that of
the universal covering group of the
diffeomorphisms of $\M$ and it is a non trivial consequence of the Lie-Rinehart
relations that it reduces to the fundamental group of $\M$.


\def \ois {{\O \in \S}}
\def \glv  {{g(\l v)}}
\def \GM  {{\G(\M)}}
\def \Gtm  {{\tilde{\G}(\M)}}
\def \tcmo {{\tilde{C}^\infty_0(\M)}}
\def \CM  {\Pi(\M)}
\def \mtre {\vspace{-3.5mm}}

\section{Lie-Rinehart $C^*$-algebra}
In this Section, we discuss how to associate to the family of  Lie-Rinehart
algebras $\lo$,   a unique ``Lie-Rinehart'' $\cst$-algebra $\am$. We adopt the
following \goodbreak
{\bf Notations}:
\begin{description} \mtre \item {$\M$} a
connected $C^\infty$ manifold of dimension $d$,
\mtre \item{$\O$} any subset of $\M$ diffeomorphic to an open sphere,
\mtre \item{$\difm$} the connected component of the identity of
the group of diffeomorphisms of $\M$,
\mtre \item{$\lm, \, \lo$} the Lie algebra of $C^\infty$ vector fields $v$ of compact
support in $\M, \O$, respectively,
\mtre \item{$\glv$}, $\l \in \Rbf$, $v \in \lm$, the associated one
parameter groups, which exist by compactness of supp\,$v$,
\mtre \item{$\gm$} the subgroup of $\difm$ generated by the $\glv$,
\mtre \item{$\gtm$} its universal covering group, which is uniquely associated to
$\lm$ (~\cite{Mi}, Theor.8.1) and  is generated by the elements of a
neighborhood of the identity in $\GM$ and therefore by the  one parameters
groups $\glv$,
\vspace{-8mm} \item{$\tilde{\G}(\O)$} the subgroup of $\Gtm$
generated by the one parameter groups $\glv$, $v \in \lo$,
\mtre \item{$\co$}
the *-algebra of $C^\infty$ complex functions on $\M$, with support in $\O$,
\mtre \item{$\cmo$} the *-algebra of  $C^\infty$ complex functions
with compact support in $\M$,
\vspace{-8mm} \item{$\tcmo$} the *-algebra generated  by $\cmo$  and the
constant functions,
\mtre \item{$\CM$} the  crossed product $\CM \eqq \tcmo \times \Gtm$.
\end{description}

The requirements 1,2 lead  to consider the *-algebra generated by $\cmo$ and
the elements $\ulv \eqq U(\glv)$, $v \in \lm$, with \be{ \a^*(x) \eqq
\bar{\a}(x), \,\,\,\forall \a \in \cmo, \,\,\,\, U(g)^* \eqq U(g^{-1}),
\,\,\,\forall g \in \Gtm,}\ee  with the Lie algebra relations between vector
fields and $\cmo$ codified by the crossed product relations, $\forall g, h \in
\Gtm$, $ \a \in \cmo$ \be{ U(g) \a(x) U(g)^{-1} = \a(g^{-1} x) \eqq \a_g(x),
\,\,\,\,U(g) U(h) = U( g h ).}\ee Thus, we are led to the
* crossed product $\CM$.

In order to impose the LR condition (1.1), since the momenta $\tv$ cannot be
bounded operators, we consider the *-algebra generated by $\CM$ and the
elements $\rv,\, v \in \lo$, playing the role of the re\-sol\-vents of the
generators of the corresponding $\ulv$, formally $\rv = (\tv - i)^{-1}$. Since
in Hilbert space representations, from the spectral representation of $\tv$,
one has (in the operator norm topology)
\be{\mbox{norm}-\lim_{\l \ra 0} [\,i (\ulv - \id)/\l
- i\, \id\,]\, \rv^2 = \rv,}\ee
this equation will be taken as the basic relation between $\rv$ and
$\ulv$, at the algebraic level.

Furthermore, the elements $\rv$ are required to satisfy the standard relations
with their adjoints \be{\rv - \rv^* = 2 i \rv \,\rv^* = 2 i \rv^* \,\rv,
\,\,\,\,\,\,\rv^* = - R_{-v}}.\ee  The Lie algebra relations obeyed by the
$\tv$ yield for the resolvents \be{ U(g)\,\rv \,U(g)^{-1} = R_{gv}, }\ee
where $gv$ is the adjoint action of $g$ on $v$.

The LR relations can be written in terms of the resolvents as
\be{R_{\a v} \, \a  -
\rv = \rav\, i(1 - \a) \,\rv - \ume  \,\rav \, \a'_v \, \rv, }\ee
$\forall \a \in
\co$,\,$\forall v \in \lo$, $\a'_v \eqq [\,v, \,\a\,] = -i \,d [U(-\l v) \a
\ulv]/d \l|_{\l = 0}$. In fact, eq.\,(2.6) is obtained  by multiplying
eq.\,(1.1) by $\rv$ and $\rav$ on the right and on the left, respectively.


In conclusion, eqs.\,(2.1-2, 2.4-6) define an abstract *-algebra $\A_0(\M)$
which incorporates the algebraic relations of the Lie-Rinehart algebras $\lo$
in terms of bounded operators. In order to make it a $\cst$-algebra we
introduce as $\cst$ norm the sup of the $\cst$ norms of $\A_0(M)$ which satisfy
eq.\,(2.3); the existence of at least one such a $\cst$ norm is guaranteed by a
(non trivial) representation of $\A_0(M)$, differentiable in the group
parameters and satisfying (1.1) (see Sect.\,3). The sup of such $\cst$ norms is
finite on $\A_0(M)$, because, for all $\cst$ norms, $||U(g)|| = 1$, $||\rv||
\leq 1$, as a consequence of eq.\,(2.4), which implies $||\rv||^2 = ||\rv^*
\rv|| \leq ||\rv||$. Moreover,
\be{||\a||  \leq \inf \{ K : |\l| > K
\Rightarrow (\a - \l)^{-1} \in \cmo \} = \sup_{x \in \M}\,|\a(x)|.}\ee
Actually, from the Schroedinger representation of $\A_0(\M)$, (see below), it
follows  that $||\a|| =  \sup_{x \in \M} |\a(x)|$ and the $\cst$-algebra
generated by $\cmo$ is $C_0^0(\M)$. The result is a unique $\cst$-algebra
$\am$, which can be considered as the ``{\em Lie-Rinehart $\cst$-algebra of
$\M$}''.

\def \ao {{\A(\O)}}

The definition of $\am$ is invariant under  $\difm$ and in fact $\Gamma_g(A)
\eqq U(g) A U(g)^{-1}$, $A \in \am$, $g \in \gtm$, defines a group of inner
automorphisms of $\am$, acting as diffeomorphism on $\cmo$ and on the vector
fields which index the resolvents. We denote by $\ao$ the subalgebra of $\am$
generated by $\a, \ulv, \rv$ with supp $\a$, supp $v \subset \O$.


\section{Regular representations of the  Lie-Rinehart $\cst$-algebra }
The notion of regular representations of the crossed product $\CM$ is well
known and amounts to the strong continuity of the one parameter  subgroups
$\ulv$, $\l \in \Rbf, v \in \lm$. By eq.\,(2.3) this property implies that $\rv
= (\tv - i )^{-1}$ on (ker$\,\rv)^{\perp}$. On the other side, if ker $\rv = \{
0\}$, eq.\,(2.3) implies strong continuity of $\ulv$ and $\rv = (\tv - i
)^{-1}$. We are thus led to

\begin{Definition} A representation $\pi$ of $\am$
is {\bf regular} if i) the representatives $\pi(\ulv)$ are strongly continuous
in $v$ in the $C^\infty$ topology of the vector fields and differentiable in
$\l$, ii) the generators $\tv = i\, d\, \pi(\ulv)/d \l|_{\l = 0}$ exist on a
common dense domain $D$ invariant under $\cmo \times \gtm$ and represent $\lm$
there, iii) ker $\pi(\rv) = \{ 0 \}$, $\forall v \in \lo$, and $\pi(\cmo) \neq
0$.
\end{Definition}

The same notion applies to representations of $\ao$. As remarked before, the
mere differentiability in $\l$ of the $\pi(\ulv)$ follows from eq.\,(2.3) and
iii). Condition iii) excludes subrepresentations with $\pi(\rv) = 0$ and
trivial representations of $\cmo$, yielding one dimensional representations of
$\am$.

\begin{Proposition} In a regular representaion $\pi$ of $\am$, the generators
$\tv$, $v \in \lm$, are essentially self-adjoint on $D$ and  satisfy $(\tv -
i)^{-1} = \pi(\rv)$, and on $D$ \be{T_{\sum_i \a_i v_i} = \ume \sum_i (\a_i
T_{v_i} + T_{v_i} \a_i), \,\,\,\,\,\forall \a_i \in  C^\infty_0(\O_i),\,\,
\,\forall v_i \in \L(\O_i).}\ee \end{Proposition}
\Pf \,\,Essential
self-adjointness on $D$ follows from invariance of $D$ under the groups
$\pi(\ulv)$, as in the proof of Stone's theorem.
\noindent
Eq.\,(2.3) implies that Range\,$(\tv \pm i) \supset \pi(R_{\pm v}) \H$, which
is dense by condition iii), ker\,$\pi(R_{\pm  v}) = \{ 0 \}$ and eq.\,(2.4).
Therefore, $\tv$ is essentially self-adjoint on $\pi(\rv)^2\,\H$. Moreover,
$\pi(\rv)$ and $(\tv - i)^{-1}$ coincide on the dense domain $\pi(\rv) \H$, and
therefore on $\H$, so that the self-adjointness domain $D(\tv)$ is $\pi(\rv)
\H$. Hence $D \subseteq \pi(\rv) \H$, $\forall v$. Eq.\,(2.6) gives
$$\pi(\rv)\,[ T_{\a v} - \ume (\a \tv + \tv \a)\,]\,\pi(R_{\a v}) = 0, \, \, \,
\,\,\,\forall \a \in \co, \,\,\forall v \in \lo$$ and therefore eq.\,(1.1)
holds on $D \cap \pi(R_{\a v}) \H = D $. Eq.\,(3.1) follows from condition ii).
 \ \ $\Box$

For the classification of the regular representations of $\am$ the following
notions are useful

\begin{Definition} A Schroedinger representation $\pi$ of $\ao$ is  a
representation in $\H_\pi = L^2(\M, d \mu)$, with $d \mu$ equivalent to the
Lebesgue measure in any coordinate system, of the following form, $\forall \psi
\in \H_\pi$
\be{(\pi(a) \psi)(x) = \a(x) \psi(x)\, , \,\,\,\,\forall \a \in
\co,}\ee
\be{ (\pi(\ulv) \psi)(x) = \psi(\glv^{-1} x) \, J(\glv,x)\,  ,
\,\,\,\,\forall v \in \co,}\ee
\be{(\pi(\rv) \psi)(x) = ((\tv - i)^{-1}
\psi)(x).}\ee
Thanks to the isometry $\psi(x) \ra [\,d \mu_2(x)/ d
\mu_1(x)\,]^{-1/2} \psi(x)$  all Schroedinger  representations $\pi$ of $\ao$
are unitarily equivalent, and therefore one may  refer to {\bf the Schroedinger
representation} $\pi_S$.\end{Definition}

\begin{Definition} Two representations $\pi_1, \,\pi_2$ of
$\am$ are locally quasi equivalent if $\pi_1(\ao) \simeq \pi_2(\ao)$, for all
$O$.

A representation $\pi$ of $\am$ is locally Schroedinger if it is locally quasi
equivalent to the Schroedinger representation $\pi_S$.
\end{Definition}

Since $\gm$ acts by inner automorphisms on $\am$, $\pi_1(\ao) \simeq
\pi_2(\ao)$ for a single $\O$ implies quasi equivalence for all $\O$. Within
the equivalence class of  locally Schroedinger representations, one may take $d
\mu(x) = d x$, $ x \in \O$, in local coordinates, so that the representation is
regular with $D = \cmo$. In this section we shall prove the following

\def \cg {{C_g}}
\def \vg {{V_g}}
\def \vh {{V_h}}

\begin{Theorem} All regular representations $\pi$ of $\am$ are locally
Schroedinger. Each $\pi$ is uniquely determined by the collection $\{
\pi(\ao)$,\,  $\O$  diffeomorphic to spheres $\}$.
\end{Theorem}
It is worthwhile to remark the relation with the representations of the crossed
product $\Pi(\M)$:

\begin{Definition} A representation $\pi$ of the crossed product $\Pi(\M)=
\tcmo \times \gtm$ is {\bf Lie-Rine\-hart (LR) regular} if it satisfies
conditions i), ii) of Definition 3.1, eq.\,(1.1) and $\pi(C_0^\infty(\M)) \neq
0$. $\pi(\Pi(\M))$ is locally Schroedinger if eqs.\,(3.2-3) hold.
\end{Definition}

\begin{Theorem} LR regular representations of the crossed product
$\Pi(\M) = \tcmo \times \gtm$ define regular representations of $\am$ and
viceversa. In particular they are locally Schroedinger and are determined by
their  restrictions to the corresponding local subalgebras.
\end{Theorem}
\Pf \,\,Given a LR regular representation $\pi$ of $\Pi(\M)$, the concrete
algebra generated by $\pi(\cifm \times \gtm)$ and the family of $R_v \eqq (T_v
- i )^{-1}$, with $T_v$ the generator of $\pi(U(\glv)$,  represents $\am$, the
LR relations, eq.\,(2.6), following from eq.\,(1.1). The converse follows from
Proposition 3.2. The last statement follows from  Theorem 3.5. \ \ $\Box$

\begin{Lemma} A regular irreducible representation
$\pi$ of $\am$ is defined  in a  separable  Hilbert space $\H_\pi$. Any such a
representation is unitarily equivalent to one with $$\H_\pi = L^2(\M, d \mu)
\times K, $$ where $d \mu$ is equivalent  to the Lebesgue measure on $\M$ (in
any coordinate system). For all vectors of $\H_\pi$, i.e. for all $L^2$
functions $\psi: \M \ra K$, the representation  is defined by
\be{\pi(\a)
\psi(x) = \a(x) \psi(x), \,\,\,\,\,\,\,\pi(U(g)) = C_g V_g, }\ee
\be{ \cg
\psi(x) \eqq \psi(g^{-1} x) \, [d \mu(g^{-1} x)/ d \mu(x)]^{1/2}, \,\,\,\,\,\vg
\psi(x) = \vg(x) \psi(x),}\ee
$\vg(x)$ a family of unitary operators in $K$,
weakly measurable in $x$, satisfying
\be{ C^{-1}_h\,V_g(x)\,C_h = V_g(hx),}\ee
\be{\vg(h x) \vh(x) = V_{gh}(x).}\ee
Two regular irreducible representations
$\pi_1$, $\pi_2$ of $\am$ are unitarily equivalent iff there exists a weakly
measurable family of unitary operators $S(x): K \ra K$, such that \be{ S(g
x)\,V_g^{(1)}(x)\,S(x)^{-1} = V_g^{(2)}(x).}\ee
\end{Lemma}
\Pf\, $\cmo$ is separable in the norm topology, (see eq.\,(2.7)); as a vector
space, $\lm$ is separable in the $C^\infty$ topology and the strong continuity
of $v \ra \ulv$, condition i), implies separability of $\pi(\gtm)$ in the
strong topology;  condition i) implies the strong continuity of $v\ra \rv$
(Theorem VIII.20 of Ref.~\cite{RS}). Hence $\pi(\am)$ is separable
in the strong topology and all the cyclic representations, in particular
the irreducible ones, are defined in a separable space.

\noindent
Since the one point compactification of $\M$, $ \dot{\M}$, is the spectrum of
the norm closure of $\tcmo$, any $\Psi \in \H_{\pi}$
defines a (Borel) measure on $\dot{\M}$ and
therefore, by separability, $\H_\pi$ can be written as \be{\H_\pi = \oplus
\sum_n L^2(\dot{\M}, d \mu_n), }\ee
with $\pi(\cmo)$ acting as multiplication
operators and $d \mu_n$ the Borel measures on $\dot{\M}$ defined by a maximal set
of vectors $\Psi_n$ giving rise to a sequence of cyclic representations of
$\tcmo$.

\noindent
Since condition iii) excludes one dimensional representations of $\am$
corresponding to the  point at infinity of $\dot{\M}$, one may replace
$\dot{\M}$ with $\M$. Hence, by defining e.g. the measure $d \nu(x) \eqq \sum_n
2^{-n} d \mu_n(x)$, one has $d \mu_n(x) = G_n^2(x) d \nu(x)$, with $G_n(x)$ $d
\nu$-measurable functions and
\be{ (\Psi, \,\Phi)_{\H_\pi} = \sum_n \int d
\nu(x) \overline{\Psi_n}(x)\, \Phi_n(x)\,G^2_n(x), \,\,\,\,\H_\pi = \int d
\nu(x)\,\H(x), }\ee
with $\H(x) \subset l^2$, dim\,$\H(x)$ a measurable
function of $x$.

\noindent
The absolute continuity of $d \nu$ with respect to the Lebesgue measure (in any
coordinate system) obviously amounts to that of $d \mu_n, \, \forall n$. Hence,
it is enough to prove  that for any Borel set $S \subset \O$ of zero Lebesgue
measure, one has $\mu_n(S) = 0$, $\forall n$. In fact, for $|\l| < \eps$,
$$\mu_n(S^\l) \eqq \int d \mu_n(x) \,\chi_S(x + \l) = (\Psi_n, \, \chi_S^\l\,
\Psi_n)$$
is a positive continuous function of $\l$, because $x \ra x + \l$, $
x \in \O$, can be obtained by the action of the groups $\ulv$, which are
strongly continuous in $\l$, by condition i). Then, by the Fubini-Tonelli theorem
$$\int d \l\, \mu_n(S^\l) = \int d \mu_n(x) \int d \l\,
\chi_S(x + \l) = 0 \, , $$
which implies $\mu_n(S) = 0$. Hence $d \nu $ is of the form $N(x) d x$, and the
function $d(x) \eqq$ cardinality of the set $\{n, \,N(x) G^2_n(x) > 0. \,x \in
\M\}$ is measurable (as a sum of the measurable functions $\theta(N(x)
G^2_n(x))$);
$d(x)$ is invariant under $ \gm $, because so are the measurable sets
$A_n \eqq \{x, d(x) = n  \}$, $ n = 0, 1 ... \infty $, since, a.e.,
$ d(x) = \mbox{dim} \, \H(x) $.
Therefore, $d(x)$ is constant, a.e. with respect to the Lebesgue measure,
since $\int d x\,\, \chi_{A_n}(x) [\a(x) - \a(x + \l)] = 0$,
$\forall \a \in \co$ and only the constant functions are orthogonal to  $\a(x)
- \a( x + \l)$, $\forall \a \in \co$, $\forall \O$. Thus $\H_\pi$ is of the form
$L^2(\M, dx) \times K$
and the first of eqs.\,(3.5) holds by construction.

\noindent
The operators $C_{\glv}$ are unitary and strongly continuous and therefore so
are the $V_g$ defined by eqs.\,(3.5), (3.6). Moreover,
$\forall \a \in \co$, $ g \in \gtm$
$$ V_g \a \psi \eqq C_g^{-1} U(g) \a \psi = C^{-1}_g \a_g  U(g) \psi = \a
C^{-1}_g U(g) \psi = \a V_g \psi,$$
and therefore  $V_g$ are decomposable
operators, i.e. they define a weakly measurable family of unitary operators
$V_g(x)$ (Theorem 7.10 of Ref.~\cite{T}).

\noindent
Eq.\,(3.7) follows from eq.\,(3.6) and eq.\,(3.8) follows from from eq.\,(2.2)
and eqs.\,(3.6), (3.7).

\noindent
The equivalence of irreducible representations of the form (3.5),(3.6) implies
dim\,$K_1 =$ dim\,$K_2$, since dim\,$K_i$ is the multiplicity of $\pi(\cmo)$.
Then, identifying $K_1 = K_2$, $S \pi_1 S^{-1} = \pi_2$ implies that $S$
commutes with $\pi_i(\cmo)$ and, therefore, by the same argument as above,  it
defines a weakly measurable family of unitary operators $S(x): K \ra
K$. \ \ $\Box$
\goodbreak

\def \ugl {U(g[\l])}
\def \ugm {U(g[\mu])}

\vspace{2mm}
For the local analysis of the representation it is convenient to
consider, for each region  $\O$, $d$ vector fields $v_i \in \l(\O')$, $\O'
\supset \O$, which define cartesian coordinates in $\O$; henceforth,  $\O$ will
be identified with the unit ball in $\Rbf^d$ and
 $\O_{\eps}$
will denote the corresponding sphere of radius $\eps$. The one parameter groups
generated by such vector fields will be denoted by $g_i(\l_i)$; then, $\forall
x \in \O$,
for $x + \l \in \O $ and $\l$ small one has
\be{ g[\l] x = x
 + \l, \,\,\,\,\,\,g[\l] \eqq \prod_{i = 1} ^d g_i(\l_i).}\ee
The diffeomorphisms $g[\l]$
depend on the choice of the defining factors $g_i[\l_i]$ and need not to
commute, even for small $\l$, since eq.\,(3.12) holds only for $x \in \O$.
However, the LR relations, eq.\,(1.1), allow to transfer the local cartesian
structure of the one parameter subgroups $g_i(\l_i)$ to the operators $\ugl$,
i.e.

\begin{Lemma} For all $\a$ with support in $\O_{1/2}$, $\l, \mu$ small,
one has
\be{ \ugl \, \ugm \, \a = U(g[\l+\mu]) \, \a.}\ee
\end{Lemma}
\Pf \,\,It suffices to prove eq.\,(3.13) for the one parameter subgroups, i.e.
$U(g_1(\l_1)) \, U(g_2(\l_2)) \, \a = U(g_2(\l_2)) \, U(g_1(\l_1)) \, \a$.
For this purpose we compute, for $\l, \mu$ small, $|\mu| < |\l_1|$,
on the common invariant domain $D$,
$$(d/d\mu) \,U(g_1(\l_1 - \mu)) \, U(g_2(\l_2)) \, U(g_1(\mu))\,\a = $$
$$= i \,U(g_1(\l_1 - \mu)) \, U(g_2(\l_2))\,(T_{g_2 v_1} - T_{v_1})
\,\a_{g_1(\mu)} \, U(g_1(\mu)) = 0,  $$
where the covariance eqs.\,(2.2), (2,5), yielding
$ U((g_2(\l_2))^{-1}\,T_{v_1}\,U(g_2(\l_2)) = T_{g_2 v_1}$,
have been used and
the last equality follows from the LR relations, eq.\,(3.1), since $g_2 v_1$
and $v_1$ coincide on the support of $\a_{g_1(\mu)}$, for $\l_1, \l_2$ small
enough. \ \ $\Box$

\begin{Lemma} For $\l,\, \mu$ small, $\forall x  \in \O_{1/2}$, one has
\be{V_{g[\l]}(g[\mu] x)\,\, V_{g[\mu]}(x) = V_{g[\l + \mu]}(x).}\ee
Moreover, a.e. in $y$,
\be{W_y(x) \eqq V_{g[x - y]}(y), \,\,\,\,x, y \in \O,}\ee
define a family  of unitary operators in $K$, weakly measurable in x,
and therefore  a unitary operator
$W_y$ in $L^2(\O, d x) \times K$, given by $W_y \psi(x) =
W_y(x) \psi(x)$; $\forall x,\,y \in \O_{1/3}$, $\l$ small they satisfy
\be{W^{-1}_y(g[\l] x) V_{g[\l]}(x) W_y(x) = \id, \,\,\,\,\,\mbox{a.e. in}
\,\,\,y  .}\ee
\end{Lemma}
\Pf \,\,Eq.\,(3.14) follows from eqs.\,(3.8) since, by eq.\, (3.13),
$$V_{g[\l]\,g[\mu]} \,\a
= C^{-1}_{g[\l]\, g[\mu]} U(g[\l]\,g[\mu])\,\a =
C^{-1}_{g[\l+\mu]}\,U(g[\l+\mu]) \,\a = V_{g[\l +\mu]} \,\a.$$
Moreover, $\forall \psi \in \H_{\pi}$,
with supp\,$\psi \subset \O$, $V_{g[\l]}(x)\,\psi(x)$
is continuous in $\l$ as an element of $L^2(\O, \,d x) \times K$ as a
consequence of Lemma 3.8 and the strong continuity of the one parameter
subgroups $g_i(\l_i)$. This implies that, $\forall \chi \in K$, $F(\l, x) \eqq
(\chi,\, V_{g[\l]}(x)\,\psi(x))_K$ is a continuous function of $\l$ in
$L^2(\O,\,d x)$.
Given a basis $e_n(x)$ in $L^2(\O,\, d x)$, $F(\l, x) = \sum
c_n(\l)\,e_n(x)$, with $c_n(\l)$ continuous in $\l$. Hence, $V_{g[\l =
x-y]}(x)$ is measurable in $x$ and $y$ and therefore, by the Fubini-Tonelli
theorem, in $x$ a.e. in $y$.
Eq.\,(3.16) follows from eq.\,(3.14) since
$$V_{g[\l]}(x)\,W_y(x) = V_{g[\l]}(g[x - y] y)\,V_{g[x - y]}(y) = V_{g[ \l + x
- y ]}(y) =$$ $$= W_y(x + \l) = W_y(g[\l] x). \ \ \Box  $$

\noindent
{\em Proof  of  Theorem 3.5}. By using the above Lemmas, $\forall \a $ with
supp\,$\a \subset \O$, $\forall U(g[\l])$, $|\l| < \eps$, $\eps = 1/8d$, a.e.
in $y$, $y \in \O_\d$, $ \d << \eps$, one has
$$ W^{-1}_y(x) \, U(g[\l]) \,W_y(x) \,
\a(x) = W^{-1}_y(x)\,C_{g[\l]}\,V_{g[\l]}(x) \, W_y(x) \, \a(x) = $$
$$ = C_{g[\l]}\,W^{-1}_y(g[\l] x)\,V_{g[\l]}(x) \, W_y(x) \, \a(x)
= C_{g[\l]} \, \a(x).$$
This implies, on $D$,
$$ -i \,T_{v_i} \a(x) = W_y(x) \, d/d\l_i \, C_{g[\l]}|_{\l =
0} \, \a(x) \, W^{-1}_y(x)$$
so that $\forall v = \sum \a_i(x)\,v_i$,  with
supp\,$\a_i \subset \O_\eps$, and therefore $\forall v \in \L(\O_\eps)$, the LR
relations (3.1) imply
\be{T_v = \sum_i T_{v_i}\,\a_i(x) + i \a'_{v_i} = W_y(x) \,
i d/d\l \, C_{\glv}|_{\l = 0} \, W_y(x)^{-1},}\ee
where the last equality follows
from the LR relations for the generators in the Schroedinger representation. The
same equation is satisfied by any extension of $W_y(x)$ to $x$ outside $\O$,
e.g. $W_y(x) = 1$, for $x \notin \O$. Eq.\,(3.16) provides therefore a unitary
equivalemce in $L^2(\M,\,d \mu) \times K$ which extends to the exponentials
and the resolvents by the essential self-adjointness  of the $T_v$ on $D$.

\noindent
In conclusion, the local properties of the unitary operators $W_y(x)$, $ x \in
\M$, imply the quasi equivalence of $\pi(\A(\O))$ in $L^2(\M, d \mu)\times K$
to the Schroedinger representation $\pi_S(\A(\O))$  in $L^2(\M, d x)$: $\forall
\psi \in L^2(\M, d \mu) \times K, \,\,\,\,A \in \A(\O)$, \be{\pi(A) \psi =
W_y\, \pi_S(A)\, W_y^{-1} \psi, \,\,\,\,\,\,\,\mbox{a.e. in}\,\, y  \in \O.}\ee
In particular, for $A = U(g)$, $\forall g \in \tilde{\G}(\O)$, $\forall x \in
\M$,  one has \be{ (\pi(U(g)) \psi)(x) = C_g W_y(g x)\,W_y(x) \psi(x),
\,\,\,\mbox{i.e.}\,\, V_g(x) = W_y(g x)\,W_y(x)^{-1}.}\ee Since, by the
regularity condition ii) of Definition 3.1, the generators satisfy $T_{v =
\sum_i v_i} = \sum_iT_{v_i}$, $v \in \lm$, by the compactness of the support of
$v$ and the essential self-adjointness on $D$, $T_v$ is uniquely determined (as
a self-adjoint operator) by the $T_{v_i}$, $v_i \in \L(\O_i)$. Hence the
representation is uniquely determined by the $\pi(\A(\O_i))$. \ \ $\Box$

\def \Vglv {V_{\glv}}
\def \gmv {g(\mu v)}
\def \loi {\L(\O_i)}
\def \wyx  {W_y(x)}


\section{Classification of the regular representations and topological effects}

For the classification of the regular representations one has to analyze the
extension of  the local characterization of the unitary operators $\Vglv(x)$,
$v \in \lo$, eq.\,(3.19), to the general case $v \in \lm$. This will be done by
following the integral curves  $\g(v, \l, x) \eqq \{ g(\mu v) x, \,0 \leq \mu
\leq \l\,\,\}$, $v \in \lm$,  patching together  the local action  of $V_{g(\l
v_i)}$, eq.\,(3.19),  $v_i \in \L(\O_i)$. The resulting characterization  of
the $\Vglv$ will depend on the equivalence class of the path $\g(v, \l)$, so
that the classification will turn out to be provided by the unitary
representations of the fundamental group $\pi_1(\M)$.

For this purpose we start with the following  preparatory Lemmas.
\begin{Lemma} The unitary operators $W_y(x)$, defined  by eq.\,(3.15),  satisfy
\newline i) $ \wyx^{-1} = W_x(y)$, \,\,\,\,a.\,e. in $(x, y)
\in \O \times \O$,
\newline ii) \, $\W(y, x) \eqq W_z(y)\,W_z(x)^{-1}$, $x,\,y,\,z \in \O$, is
independent
of $z$ a.e. in $\O$  and therefore it is well defined a.e. in $x \in \O$ and
a.e. in $y \in \O$ and satisfies  \be{\W(y, x) \,\W(x,\,z) = \W(y,\,z).}\ee
\end{Lemma}
\Pf\,\,Property i) follows trivially from the definition (3.15) and eq.\,(3.8).
The independence of $z$ in ii) follows from i) and eq.\,(3.8), and implies
eq.\,(4.1) $\forall x, y \in {\cal O}\diagdown I$, $I$ a set of zero measure.
 \ \ $\Box$

\begin{Lemma} Let $\g(y, x)$ be a $C^\infty$ curve starting at $x$ and ending
at $y$, and   $\g(y, x) = \g(y, x_n) \circ \g(x_n, x_{n-1}) \circ
...\g(x_1,x)$
be a partition such that $\g(x_{i+1}, x_i) \in \O_i$, then
\be{ \W(y, x, \g(y, x))  \eqq \W(y, z_n)\,\W(z_n, z_{n-1})...\W(z_1, x),}\ee
is independent of the points
$z_i \in \O_{i-1} \cap \O_i$ and of the partition chosen in
eq.\,(4.2); it depends only on the (homotopic) equivalence class $[\g(y, x)]$,
i.e. $\W(y, x, \g(y, x)) = \W(y, x, [\g(y, x)])$. Furthermore, $\forall x, y,
z, \in \M $
\be {\W(y, x, [\g(y, x)])\,\W(x, z, [\g(x, z)]) = \W(y, z, [\g(y,
x) \circ \g(x, z)]).}\ee
The operators $\W(x, x, [\g]): K \ra K$ are well
defined a.e. in $x \in \M$ by ii) of Lemma 4.1 and  define unitary
representations of $\pi_1(\M)$, which are unitarily equivalent for all
$x \in \M$.
\end{Lemma}
\Pf\,\,The construction (4.2) is independent of the addition and displacement
of an intermediate point as a consequence of eq.\,(4.1). The composition law
(4.3) follows from eq.\,(4.2) with $x$ as intermediate point. Given $x$ and
$y$, an equivalence relation between the closed curves $\g(x,x) $, and
$\g(y,y)$ is given by  $[\g(y,y)] = [\bar{\g}(y, x)\circ\,\g(x, x)
\circ\,\bar{\g}(x, y)^{-1}]$, with $\bar{\g}(x, y)$ a fixed curve; then the
unitary equivalence of the representations of $\pium$ is given by $\W(y, x,
[\bar{\g}(y,x)])$. \ \ $\Box$

\def \gvl {\g(v, \l)}
\def \\glvt {g(\l \tilde{v})}

\begin{Lemma} The operators \be{ \W_{\glv}(x) \eqq \W(\glv x, x,
[\g(v)]),\,\,\,\,v \in \lm, }\ee with $\g(v) \eqq\g(v, \glv x, x)$  the
integral curve of $v$ starting at $x$ and ending at $\glv x$, define, for fixed
$v,\, \l$,\, a family of unitary operators in $K$, weakly measurable in $x$,
and therefore unitary operators in $L^2(\M, d x) \times K$, with the following
properties
\newline i) \,\,$\forall v  \in \lm$, $ h \in \gtm$,
$$\W_{\glv}(x)\,C_{h} = C_{h} \W_{\glv}(h x)$$
ii)\,\,$ \U(\glv) \eqq
C_{\glv}\,\W_\glv(x)$ form a one parameter  group in $\l$
\newline iii)\,\,$\forall v \in  \lm$, for all $\O$,  $ \W_\glv(x)
= \W_{g(\l \tilde{v})}(x) = V_{g(\l \tilde{v})}(x)$, \,\,\,$\forall x \in
\O_{1/2}$, $\l$ small enough, $\forall \tilde{v} \in \lo$ with $\tilde{v}  = v
$ in $\O_{3/4}$.
\end{Lemma}
\Pf \, \, Property i) follows from the definition and properties of
the $\W$ as a family
of unitary operators in $L^2(\M, d x) \times K$. The group properties follow
from those of the $\W$, eq.\,(4.3), and from property i).
Property iii) follows from  the definition of the $\W$ and eq.\,(3.19):
$$\W_\glv(x) = W_{g(\l \tilde{v})}(x) = W_y(g(\l \tilde{v}) x)\,W_y(x)^{-1} =
V_{g(\l \tilde{v})}(x). \ \ \Box $$

The results of the above Lemmas allow for an
extension of eq.\,(3.19) to $g \in \gtm$.

\begin{Proposition} For any $\glv \in \gtm$ one has
\be{ (\pi(U(\glv)) \psi)(x)
= C_g\,\W(\glv x, x, [\g(v)]) \psi(x),}\ee
i.e. $V_g(x) = \W(\glv x, x,[\g(v)])$.
\end{Proposition} \Pf\,\, We have to prove that
$\pi(U(\glv))  = \U(\glv)  \,\,\forall \,v \in \lm$.
In fact, both $\pi (U)$ and $\U$ are one parameter groups of
strongly continuous unitary operators.
The generator of $\U(\glv)$ exists on $D$ and coincides with $T_v$ there.
In fact, chosen $\a$ with compact support and $\a = 1$ on supp($v$),
$$\pi(U(\glv)) \, (1- \a) = 1 - \a = \U(\glv) \, (1 - \a), $$
so that  both generators vanish when multiplied by $1 - \a$;
by choosing $\a_i(x) \in C^\infty(\O_i)$, $\sum_i \a_i(x) = \a$,
$\tilde{v}_i = v$ in $\O_i(1 + \eps)$,
$\l $ small, one has on $D$, by Lemma 4.3, iii),
$$ \U(\glv) \sum_i \a_i(x) =
\sum_i  C_{g(\l \tilde{v}_i)} V_{g(\l \tilde{v}_i)}(x) \,\a_i(x) = \sum_iU(g(\l
\tilde{v}_i)) \, \a_i(x) $$
and therefore
$$ (d/ d \l) \, \U(g(\l v))|_{\l =
0}\, \a (x) = - i \sum_i T_{\tilde{v}_i} \a_i(x).$$
The LR relations (Proposition 3.2), equivalently the locally Schroedinger
property, and the
support properties of $v_i$ give $$ \sum_i T_{\tilde{v}_i} \a_i = \sum_i
(T_{\a_i \tilde{v}_i} + \ume \,[\,\tilde{v}_i, \,\a_i\,]) = \sum_i T_{\a_i v} +
\ume [\,v, \,\sum_i \a_i\,] = T_v = T_v \, \a .  \ \ \Box  $$

\begin{Theorem}
Modulo unitary equivalence a regular irreducible  representation of $\am$ is
characterized by the unitary irreducible representation of $\pi_1(\M)$, defined
by Lemma 4.2.
\end{Theorem}
\Pf\,\,Given two regular irreducible representations, $\pi_1, \pi_2$, of $\am$,
by Lemma 3.8
they are of the form (3.5), (3.6) in $L^2(\M, d \mu) \times K_i$, $i = 1, 2$,
and by Proposition 4.4 the corresponding $U_i(g)$ are determined by the operators
$\W(y, x, [\g])$ of Lemma 4.2. By Lemma 3.8 the unitary equivalence of the two
representations amounts to the existence of  unitary operators $ S(x)$ with
\be{ S(y) \,\W_1(y, x, [\g])\,S(x)^{-1} = \W_2(y, x, [\g])}\ee
and this
trivially implies the unitary equivalence of the corresponding representations
of $\pium$.

\def \xo {x_0}

\noindent
On the other hand, the unitary equivalence of the corresponding unitary
representations of $\pium$ reads
\be{\W_1(z, z, [\g]) = V_z^{-1} \,\W_2(z, z,
[\g])\,V_z, \,\,\,\,V_z : K_1 \ra K_2};\ee
then, a.e. in $z$, the operators
$$ S(x) \eqq \W_2(x, z, [\g])\, V_z\,\W_1(x, z, [\g])^{-1}$$
are independent of $\g$ and define unitary operators
$ S: L^2(\M, d \mu) \times K_1 \ra L^2(\M, d\mu) \times K_2 $.
In fact, given $x, \xo$ and a fixed curve $\bar{\g}(x, \xo)$,
any $\g(x, \xo)$ may be uniquely written as $[\g(x, \xo)] = [ \bar{\g}(x, \xo)
\circ \g_0(\xo, \xo)]$, $ \g_0 = \bar{\g}^{-1} \circ \g$; then
$$\W_i(x, \xo,
[\g]) = \W_i(x, \xo, [\bar{\g}])\,\W_i(\xo, \xo, [\g_0])$$ and by eq.\,(4.7)
$$\W_2(x, \xo, [\g])\,V_{\xo} \, \W_1(x, \xo, [\g])^{-1} =
  \W_2(x, \xo, [\bar{\g}]) \, V_{\xo} \, \W_1(x, \xo, [\bar{\g}])^{-1}, $$
i.e. $ S(x) $ is independent of $\g$. Since
$$\W_i(y, \xo, [\bar{\g}(y, \xo)])^{-1} \, \W_i(y, x, [\g(y,
x)])\,\W_i(x, \xo, [\bar{\g}(x, \xo)]) =$$
$$= \W_i(\xo, \xo,
[\bar{\g}(\xo, y) \circ \g(y, x) \circ \bar{\g}(x, \xo)]$$
and $V_{\xo}$ intertwines
between the right hand sides of the above equation, $i= 1,2$, one gets
eq.\,(4.6). \ \ $\Box$

\def \mt  {\tilde{\M}}
\def \xg   {(x, [\g])}
\def \ptxg  {\tilde{\psi}\xg}
\def \xt {\tilde{x}}
\def \xgx {(x, [\g_x])}
\def \lmk {L^2(\mt, K, R)}
\def \tpsi {\tilde{\psi}}

\vspace{2mm}

The regular irreducible representations of $\am$ can be given a more
explicit form, which also  exhibits their complete characterization in terms of
representations of $\pium$. By exploiting the results of Proposition 4.4, they
can be identified with Schroedinger representations  on multivalued wave
functions on $\M$, namely on wave functions on $\mt$, the universal covering
space of $\M$.

The following notions and notations are useful: the points of $\mt$ will be
denoted by the pairs $\xg$, with $\g$ a $C^\infty$ curve in $ \M $
starting at a fixed point $\xo$ and ending at $x$;
the group $\gtm$ acts naturally on $\mt$ as
$\glv \xt = \glv \xg = (\glv x, [\g(v)) \circ \g])$, $\g(v)$ as in eq.\,(4.4);
a regular immersion $x \ra \xt$ of $\M$ into $\mt$ is given by $x \ra \xgx$,
with $x \ra \g_x$ a family of curves depending continuously on $x$ in an open
subset of $\M$ with complement of zero measure.

For any given unitary representation $R$ of $\pium$ in a Hilbert
space $K$, $L^2(\mt, K, R)$ will denote the space of locally $L^2$ wave
functions $\tpsi: \mt  \ra K$ satisfying
\be{\tpsi(x, [\g \circ \g_0]) =
R([\g_0])^{-1} \tpsi\xg,}\ee
with norm
\be{|| \tpsi||^2 \eqq ||
\tpsi||^2_{\lmk} = \int_{x \in \M} d \mu(x)\,||\psi\xgx||^2_K,}\ee
where $d\mu(x)$ is any measure absolutely continuous with respect to the Lebesgue
measure $d x$ in any coordinate system. The above Hilbert norm is independent
of the choice of the family $\{\g_x\}$ since $R$ is unitary, and different
choices of $d \mu(x)$ lead to equivalent constructions.

\begin{Theorem} Any regular irreducible representation $\pi$ of $\am$ is
unitarily equivalent to the representation $\tilde{\pi}$ in $\lmk$, with $R$
the unitary representation of $\pium$ associated to $\pi$ by Lemma 4.2, defined
by \be{\tilde{\pi}(U(\glv)) \tpsi(\xt) = \tpsi(\glv^{-1} \xt) \,J(\glv, x),
\,\,\,\,\, \a \, \tpsi\xg = \a(x)\,\tpsi\xg.}\ee Conversely, any unitary
irreducible representation $R$ of $\pium$ defines a regular irreducible
representation of $\am$, given by eqs.\,(4.8-10).
\end{Theorem}
\Pf\,\, The unitary equivalence is given by the isometry
\be{T: L^2(\M, d \mu)
\times K \ni \psi(x) \ra \tpsi\xg = \W(x, \xo, [\g])^{-1} \,\psi(x),}\ee
with
$\W$ defined by eq.\,(4.2). In fact, by eq.\,(4.3)
$$ \tpsi(x, [\g \circ \g_0])
= \W(x, \xo, [\g \circ \g_0])^{-1} \, \psi(x) = \W(\xo, \xo, [\g_0])^{-1}
\, \tpsi\xg,$$ i.e. eq.\,(4.8) holds and the unitarity of $T$ follows from that of
$\W$ and eq.\,(4.9).

\def \tpi {\tilde{\pi}}

\noindent
Furthermore, $T$ intertwines between $\pi$ and $\tpi$, since by Lemma 4.3
$$C_g \,\, \W(\glv x, x, [\g(v, \glv x, x)]) = \W(x, \glv^{-1} x, [\g(v, x,
\glv^{-1} x)]) \, \, C_g, $$
$$ \W(x, \xo, [\g])^{-1}\,\W(x, \glv^{-1} x, [\g(v, x,
\glv^{-1} x)]) =$$ $$= \W(\glv^{-1} x, \xo, [\g(v, \glv^{-1} x, x) \circ
\g])^{-1}, $$
and therefore, by Proposition 4.4
$$(T \, \, \pi(U(\glv))\psi) \, \xg =
\tpsi(\glv^{-1} x, [\g(v) \circ \g]) = \tpi(U(\glv) \, \tpsi \xg.$$
The unitary equivalence obviously extends to the resolvents.

\def \cmkr {C^\infty_0(\mt, K, R)}

\noindent
Conversely, given a unitary irreducible representation $R$ of $\pium$ in a
Hilbert space $K$, eqs.\,(4.8-4.10) define a LR-regular representation of the
crossed product $\tcmo \times \gtm$. In fact, the space $C^\infty_0(\mt, K, R)$
of $K$-valued functions of compact support on $\M$ which are strongly
infinitely differentiable and satisfy eq.\,(4.8) is dense in $\lmk$, since it
contains the spaces  defined by the  extension  of   $\co$ through eq.\,(4.8)
for all $\O \subset \M$. $\cmkr$ is invariant under $C^\infty_0(\M) \times
\gtm$ and gives a regularity domain $D$, since the strong differentiability of
$\pi(U(\glv))$ on $D$ follows by a dominated convergence argument; the strong
continuity with respect to $v \in \lm$ follows similarly. The locally
Schroedinger property follows from eq.\,(4.10), since there is no dependence on
$\g$ for $x \in \O$, and implies eq.\,(3.1) and therefore LR regularity. By the
first part of Theorem 3.7 one gets a regular representation of $\am$. \ \ $\Box$

\end{document}